\documentclass{article}[12pt]
\usepackage{amsmath}
\usepackage{epsfig}
\usepackage{latexsym}
\usepackage{float}
\usepackage{amssymb}
\usepackage{graphicx}

\begin{document}

\title{Phase separation kinetics in compressible polymer
solutions: Computer simulation of the early stages}
\author{Peter Virnau$^{1}$, Marcus M\"uller$^{1}$, Luis Gonz\'alez MacDowell$^{2}$,
\\and Kurt Binder$^{1}$\\
\\
$^{1}$Institut f\"ur Physik, Johannes Gutenberg--Universit\"at Mainz,\\
Staudinger Weg 7, 55099 Mainz, Germany\\
$^{2}$Dpto. de Quimica Fisica, Facultad de Cc. Qumicas, \\
Universidad Complutense, 28040 Madrid, Spain}
\maketitle

\begin{abstract}
A coarse-grained model for solutions of polymers in supercritical
fluids is introduced and applied to the system of hexadecane and
carbon dioxide as a representative example. Fitting parameters of
the model to the gas-liquid critical point properties of the pure
systems, and allowing for a suitably chosen parameter that
describes the deviation from the Lorentz-Berthelot mixing rule,
we model the liquid-gas and fluid-fluid unmixing transitions of this system
over a wide range of temperatures and pressures
in reasonable agreement with experiment. Interfaces between the
polymer-rich phase and the gas can be studied both at temperatures above and below
the end point of the triple line where liquid and vapor carbon dioxide and the
polymer rich phase coexist. In the first case interfacial adsorption of fluid
carbon dioxide can be demonstrated. Our model can also be used to
simulate quenches from the one-phase to the two-phase
region. A short animation and a series of snapshots help to visualize the early stages of bubble
nucleation and spinodal decomposition. Furthermore we discuss deviations
from classical nucleation theory for small nuclei.
\\
\\
PACS: 64.60.Qb, 82.20.Wt, 05.70.Fh, 61.41.+e
\end{abstract}
\newpage

\section{Introduction}
\begin{figure}
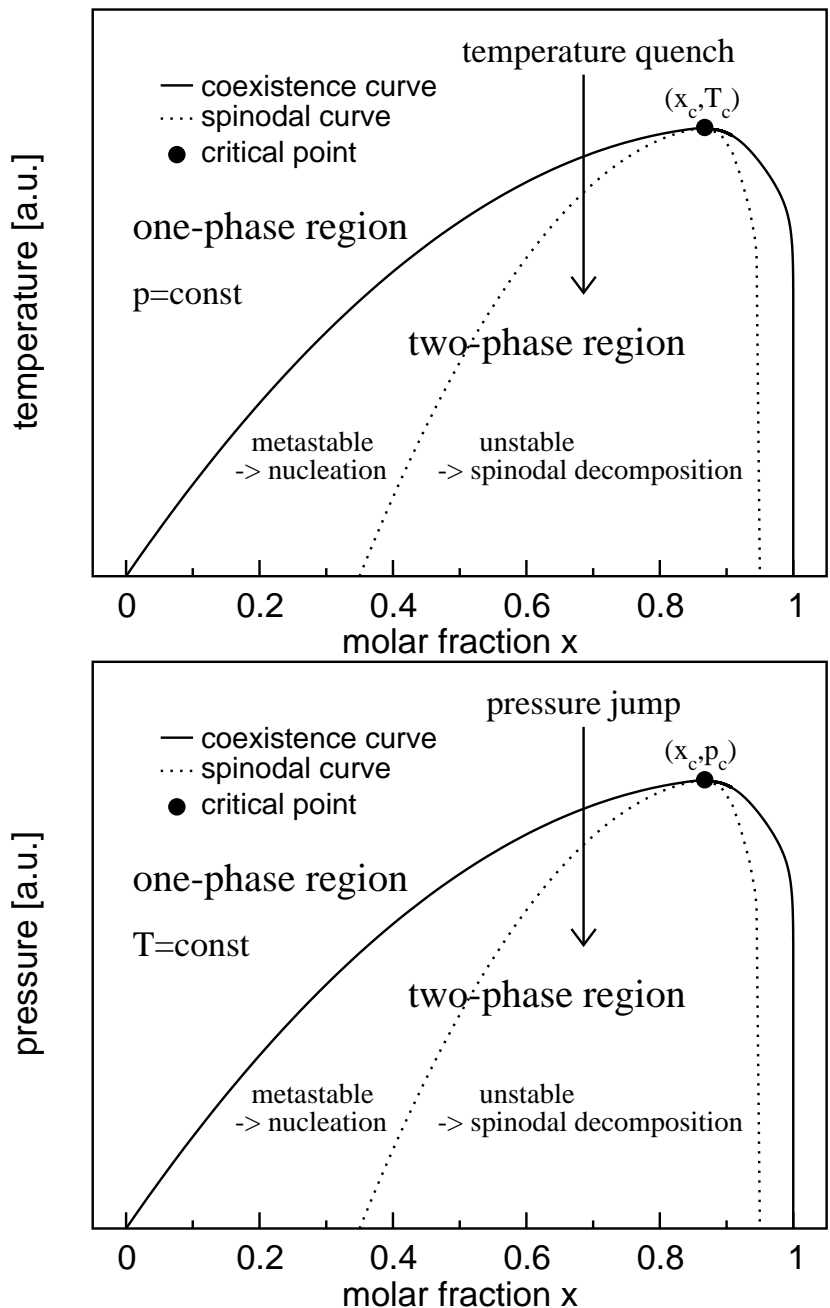

\begin{center}
\epsfig{file=Tx_schematic.eps,width=0.90\linewidth,clip=}
\epsfig{file=px_schematic.eps,width=0.90\linewidth,clip=}
\caption{\label{fig1}a) Schematic phase diagram of
an incompressible polymer solution. Temperature $T$ and molar
fraction $x$ of the solvent are variables, pressure p is constant.
The position of the critical point is shown by $\bullet$.
We indicate how a sudden
decrease of temperature from a state in the one phase region to a
state in the two-phase region is carried out. Note that the molar
fraction $x_c$ of the critical point tends to unity when the chain
length $N$ of the macromolecule gets large. b) Schematic
isothermal slice of a phase diagram of a compressible polymer
solution, using pressure $p$ and molar fraction $x$ of the solvent
as variables. We indicate how a pressure jump experiment could
be performed. In both cases a) and b) it is assumed that the state
after the jump is in the unstable region of the phase diagram, i.e.,
underneath the spinodal curve. If the quench ends in between the
spinodal curve and the coexistence curve, phase seperation will start by
homogeneous nucleation.}
\end{center}
\end{figure}

\begin{figure}
\begin{center}
\epsfig{file=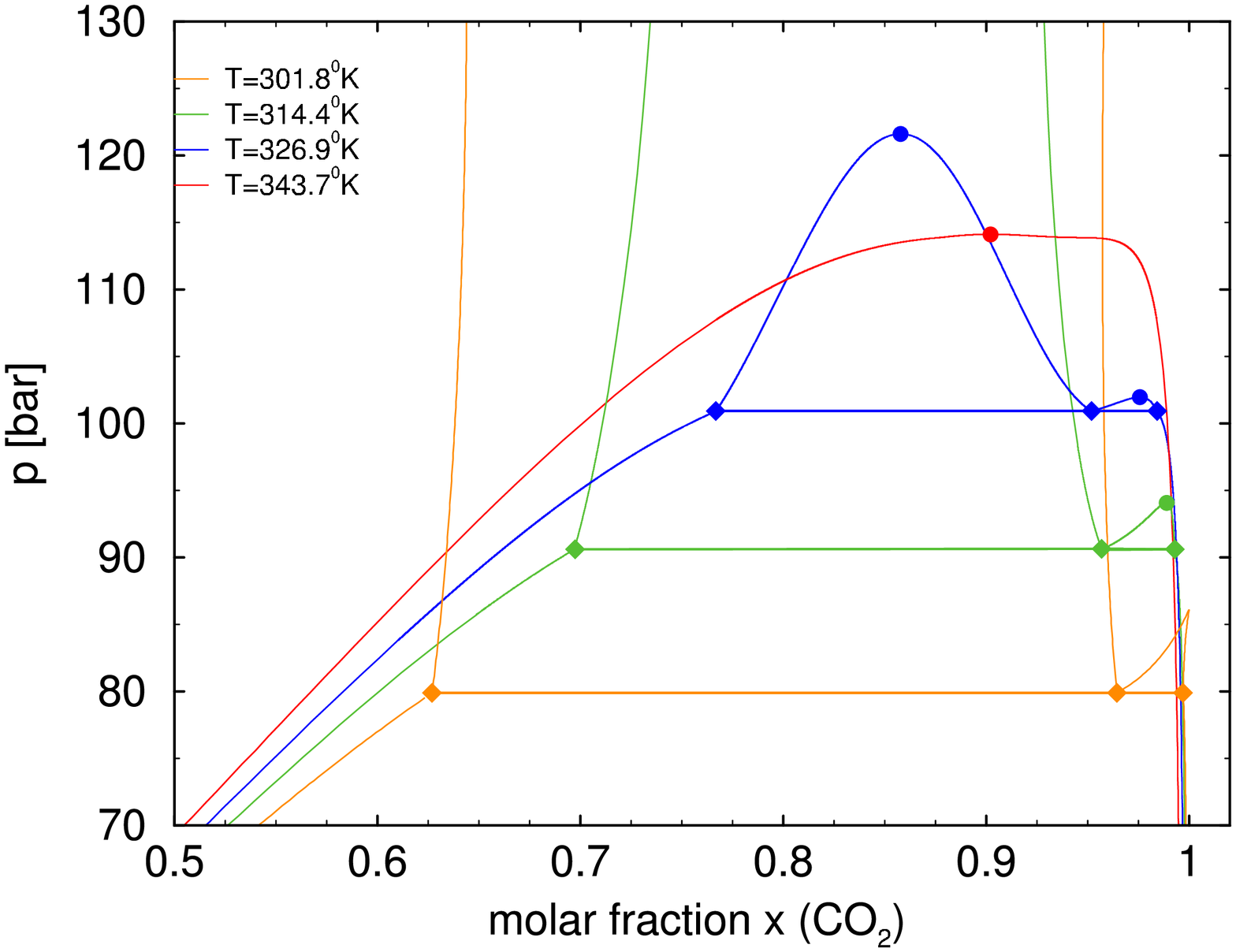,width=0.70\linewidth,angle=270,clip=}
\caption{\label{fig2} Isotherms of a simple model-system obtained from self-consistent field calculations \cite{11}. 
At T=343.7 K the phase diagram is like the schematic diagram in Fig. \ref{fig1}b. However,
at temperatures close to the critical point of pure CO$_{2}$ we observe 
liquid-liquid immiscibility for a type III system. A three phase line and a second critical point occur.}
\end{center}
\end{figure}

The phase behavior of polymer-solvent systems has important
application in the industry for the production and processing of
many kinds of plastic materials \cite{1,2,3,4,5,6}. An example is
the formation of solid polystyrene foams \cite{3,4,5,6}. In
addition, both the understanding of the equilibrium phase diagram of
these systems and the kinetic mechanism of phase separation are
challenging problems of statistical mechanics. On the one hand,
variation of the molecular weight of the polymer (i.e. the ``chain
length'' of the flexible linear macromolecule, N) offers a control
parameter that leaves intermolecular forces invariant, and hence
allows a more stringent test of the theory than would be possible
for small molecule systems. On the other hand, it is crucial to
take the compressibility of the system fully into account, since
for practical reasons one normally wishes to use supercritical
fluids (e.g.,~CO$_2$) as a solvent \cite{1,2,3,4,5,6}. Then
small changes of the pressure result in a large variation of the
solvent density, and this fact obviously facilitates applications.
While for an incompressible polymer solution (as it is modelled,
for instance, by the well-known Flory-Huggins lattice theory
\cite{7,8,9,10}) the early stages of phase separation can only be
studied if one realizes a rapid temperature quench from a state in
the one phase region into the miscibility gap (cf.
Fig.~\ref{fig1}a), for compressible polymer solutions it is
possible to induce phase separation by an (experimentally much
more convenient) pressure jump (cf. Fig.~\ref{fig1}b). Of course,
in Fig.~\ref{fig1} we have made the tacit assumption that
including the pressure does not change the qualitative form of the
phase diagram, i.e., the point ($T_c,x_c$) in $(T,x,p)$-space
becomes a line $(T_c(p), x_c(p))$ or, expressed in the appropriate
inverse functions, $(p_c(T), x_c(T))$, respectively
(Fig.~\ref{fig1}b). In reality, the situation is more complicated,
since one must treat both the densities of CO$_2$ and hexadecane 
as two coupled
order parameters, which both play a crucial role in the possible
phase transitions (remember that for the pure solvent at its
critical point gas-liquid phase separation starts). As a result,
it can happen that both gas-liquid and
liquid-liquid phase separation are present in the phase
diagram, leading to the occurrence of a triple line
(Fig.~\ref{fig2}) \cite{11}. As will be discussed below, the
presence of the triple line has a profound influence on the
interface properties and hence the nucleation behavior in the
system.

The present paper is devoted to a study of the early stages of phase
separation kinetics in such systems by computer simulation. We use
a coarse-grained model for mixtures of hexadecane and
carbondioxide as an archetypical example. The reason for the
choice of this particular system is the fact that recently
extensive experiments on nucleation in this system have been
performed \cite{12}. In Sec.~2, we shall briefly describe the
model and the simulation techniques. Sec.~3 presents the relevant
static equilibrium properties, in particular, the phase diagram and
the properties of planar interfaces separating coexisting phases.
Sec.~4 describes simulations of quenching experiments under
different conditions, that lead either to nucleation and growth or
spinodal decomposition. Sec.~5 summarizes our conclusions.

\section{Model and Simulation Techniques}
A fully atomistic simulation of mixtures of hexadecane (C$_{16}$
H$_{34}$) and carbondioxide (CO$_2$), which aims at both establishing
phase diagrams as a function of the three variables $T,p$ and
$x$, and the study of phase transition kinetics under
various conditions, would be an extremely formidable problem. The
potentials for the length of the covalent bonds in these molecules
(as well as the potentials for bond angles) are very stiff. Hence, 
an extremely short time step (of the order of about 1 fs) would be required
in a Molecular Dynamics (MD) simulation. In a corresponding Monte Carlo (MC) simulation
only very small random displacements of atoms would be admissible. Thus, for liquid
alkanes it is a quite common and well-established practice to
integrate CH$_2$-groups (as well as the CH$_3$ end groups) into
``united atoms'' \cite{13}, and also to work with a fixed C-C bond
length. These simplifications already reduce the necessary
computer time by a factor of 100. However, we have estimated that
even for such a simplified model computer resources as they are
available today still are not yet sufficient. Thus it was chosen
to simplify the model even further, representing CO$_2$ by a
single pseudo-atom, and representing C$_{16}$ H$_{34}$ by a
flexible chain of 5 subsequent effective segments, each of which
then contains roughly 3 successive C-C bonds. All such effective
monomeric units interact with a truncated and shifted Lennard
Jones (LJ) potential,
\begin{equation} \label{eq1}
V_{LJ}(r)= 4 \varepsilon_{\textrm{hh}} [(\frac {\sigma
_{\textrm{hh}}}{r})^{12} - (\frac {\sigma
_\textrm{{hh}}}{r})^{6}+\frac{127}{16384} ],\hspace{0.5cm} r \leq r_c=2\cdot2^{1/6}
\sigma \; ,
\end{equation}
while $V_{LJ}(r>r_c)=0$. Subsequent effective monomers along a
chain are in addition exposed to a finitely extensible nonlinear
elastic (FENE) potential \cite{14}
\begin{equation} \label{eq2}
V_{\textrm{FENE}}(r) = -33.75 \epsilon_{hh} \ln [1-(\frac{r}{1.5
\sigma_{\textrm{hh}}})^2 ]
\end{equation}
The constants in this potential are chosen such that the most
favorable distance between bonded neighbors is 0.96
$\sigma_{\textrm{hh}}$, while the preferred distance between
non-bonded effective monomers is $2^{1/6} \,  \sigma_\textrm{hh}
\approx 1.12 \, \sigma_\textrm{hh}$. This mismatch is desirable
to prevent crystallization at high densities \cite{15},
which is appropriate for glass-forming polymers \cite{3,4,5,6}.

The parameters $\sigma_{\textrm{hh}}$, $\varepsilon_{\textrm{hh}}$
set the scales for energy and length of our hexadecane model.
Since we wish that our model fits the thermodynamic properties of
this material in the fluid phase as faithfully as possible, we
have adjusted them such that the critical temperature $T_c=723$~K
and the critical density $\rho_c =0.21$~g/cm$^3$ of hexadecane
are correctly reproduced. In simulations, a critical point can be computed 
by a finite size scaling study of the liquid-gas transition 
along the lines of the techniques proposed by Wilding et al. \cite{16}.
A comparison of critical temperature and density in Lennard-Jones units with
the experimental values yields
$\varepsilon_{\textrm{hh}}=5.787 \cdot 10^{-21}$ J,
$\sigma_{\textrm{hh}}=4.523 \cdot 10^{-10}$m \cite{17,18}.

Since the description of hexadecane is thus already reduced to a
crudely coarse-grained model, it would not make sense to keep all
atomistic detail for carbon dioxide. Thus the CO$_2$ molecule
is also coarse-grained into a point particle, and we require an
interaction potential between CO$_2$ molecules of exactly the same
LJ form as in Eq.~(\ref{eq1}), but with parameters
$\varepsilon_{\textrm{cc}}$, $\sigma_{\textrm{cc}}$. Requiring
once more that the critical temperature and density of real CO$_2$
are correctly reproduced we obtained \cite{17}
$\varepsilon_{\textrm{cc}}=4.201 \cdot 10^{-21}$~J, and
$\sigma_{\textrm{cc}} = 3.693 \cdot 10^{-10}$~m. It is clear that
this procedure ignores some physical effects, e.g., the fact that
the CO$_2$ molecules carry electrical multipole moments is
completely neglected. Nevertheless our model not only describes
the gas-liquid coexistence curve of hexadecane and CO$_2$ over a reasonable
temperature range \cite{17}, but also other data (like the
critical pressure, or the temperature dependence of the surface
tension near $T_c$ \cite{18}).

\begin{figure}
\begin{center} 
\epsfig{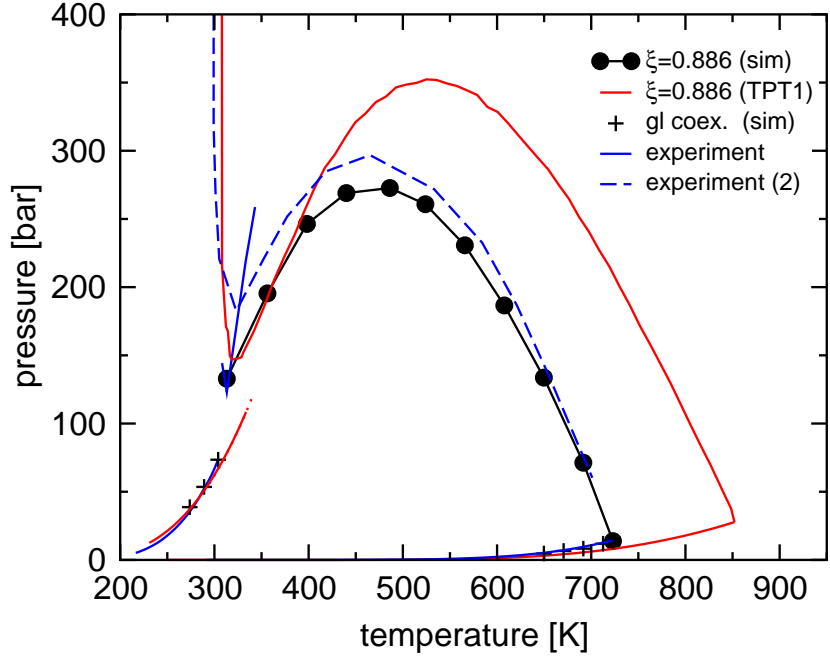}
\caption{\label{fig3} Projection of the hexadecane-CO$_2$
phase diagram (p,T,x) onto the pressure-temperature plane.
Black lines, + and $\bullet$: simulations (for $\xi=0.886$), red lines: 
analytic calculations from a thermodynamic pertubation theory (TPT1) based on the same model \cite{23},
blue lines: experimental data. Gas-liquid coexistence lines are from \cite{40,41}, the two critical lines are 
taken from \cite{38} and \cite{22,42} (dashed blue) and differ considerably.
Experiments are in qualitative agreement with both simulations and analytic calculations.
Three features can be identified: Gas-liquid coexistence lines (+) of pure CO$_2$
and hexadecane both end at the corresponding critical point (at T=304 K and T=723 K respectively).
A line of critical points ($\bullet$) emerges from the critical point of pure hexadecane and
gradually changes its composition from gas-liquid hexadecane to liquid CO$_2$ - liquid hexadecane.
The red dotted three-phase line (TPT1) lies slightly below the corresponding CO$_2$ coexistence curve, 
but only above the critical point of CO$_2$ it is distinguished in this plot. The critical point 
of CO$_{2}$ and the the end point of the three-phase line are connected by another short critical line
which cannot be distinguished, too.}
\end{center}
\end{figure}

Of course, the choice of interaction parameters between CO$_2$ and
hexadecane is more subtle. We model the interaction between the
pseudoatoms representing CO$_2$ and the pseudoatoms representing
three subsequent CH$_2$ groups again by a LJ potential of the form
Eq.~(\ref{eq1}), but now with parameters
$\varepsilon_{\textrm{hc}}$, $\sigma_{\textrm{hc}}$. It then
remains to find an optimal choice for these parameters. The
simplest choice would be the well-known Lorentz-Berthelot mixing
rule \cite{19}
\begin{equation} \label{eq3}
\varepsilon_{\textrm{hc}} = \sqrt {\varepsilon_{\textrm{hh}} \,
\varepsilon_{\textrm{cc}}} \, , \quad
\sigma_{\textrm{hc}}=(\sigma_{\textrm{hh}} + \sigma_{\textrm{cc}})
/2 \quad .
\end{equation}
However, when one tries the choice Eq.~(\ref{eq3}) the resulting
phase diagram of our model system would be of type I in the
classification scheme of Van Konynenburg and Scott \cite{20,21}
(i.e., the critical points of pure hexadecane and CO$_2$ are connected
by a critical line of the mixture system. Liquid-liquid 
phase seperation does not exist and thus there is no three phase coexistence). 
However, experimentally it is known \cite{22}
that alkane -- CO$_2$ mixtures exhibit a type I -- phase behavior
only for very short alkanes, while for hexadecane -- CO$_2$
mixtures the phase diagram is of type III. Instead of a connecting 
line we rather observe a
topology as shown in Fig.~\ref{fig3} \cite{17,23}. In the (p,T) projection 
of the phase diagram, a critical line emerges from the critical point of
hexadecane and does not end at the critical point of CO$_2$.
In addition we observe liquid-liquid immisciblity and a three-phase line. 
When we want to obtain this behavior from our coarse-grained model we
must allow for a deviation from the first equation of
Eq.~(\ref{eq3}), assuming rather
\begin{equation} \label{eq4}
\varepsilon_{\textrm{hc}} = \xi \sqrt{\varepsilon_{\textrm{hh}} , 
\varepsilon_{\textrm{cc}}} \, , \quad \xi < 1 \quad .
\end{equation}
It turns out that a suitable choice for the parameter $\xi$
which characterizes the deviation from the Lorentz-Berthelot mixing
rule is \cite{17,18,23} $\xi=0.886$. Eq.~(\ref{eq4}) with this
choice of $\xi$ was in fact used to compute the phase diagram
shown in Fig.~\ref{fig3}. However,
experimental measurements of the critical line vary considerably
and a small modification in $\xi$ rises or lowers the
critical line. Details about how such phase diagrams
are in fact estimated from the simulation are given elsewhere
\cite{18}.

For a study of the phase diagram (Fig.~\ref{fig3}) as well as for
a study of the interfacial free energy between coexisting gas and
liquid phases \cite{17,18} the grandcanonical ensemble is used,
where volume $V$, temperature $T$, and the chemical potentials
$\mu_{\textrm{c}}$, $\mu_{\textrm{h}}$ are fixed. 
Both the particle numbers $N_{\textrm{c}}$ and
$N_{\textrm{h}}$ of CO$_2$ and hexadecane and the pressure $p$ are
then (fluctuating) observables ``measured'' in the simulation
\cite{18}. 
Fig.~\ref{fig4} shows an isothermal slice through the phase diagram at
$T=486$ K.
This temperature is higher than the temperature of the critical end point
where the critical line ending at the critical
point of CO$_2$ and the line of triple points ($p_{\textrm{trip}}
(T)$) in the ($p,T$) phase diagram (Fig.~\ref{fig3}) meet.
Therefore unlike Fig.~\ref{fig2} there is no triple point in the
phase diagram, and the qualitative features are the same as in
Fig.~\ref{fig1}b. At this temperature we shall examine the
kinetics of the phase separation in Sec.4.

In order to observe the kinetics of bubble
nucleation in the vicinity of the binodal we control the
undersaturation of the system by fixing the chemical potential of
both species. As a starting configuration we use a homogeneous
state equilibrated at a higher temperature. This situation
corresponds to a simulation cell in contact with a much larger
reservoir which is held at constant undersaturation. As the number
of particles in the simulation cell is not conserved, and the
particles do not move according to a realistic dynamics, we do not
obtain information about the time scale of bubble nucleation. The
kinetics of phase separation in the vicinity of the binodal is,
however, chiefly determined by the free energy barrier the system
encounters on its path towards the equilibrium state. Hence, we
expect the relaxation path to be similar to a simulation with
realistic dynamics. The most natural choice would be to apply
Molecular Dynamics methods \cite{24,25}. However, if the number of 
particles was conserved we would need prohibitively large 
simulation cells for the undersaturation not
to decrease substantially even in the very early stages of
nucleation. The simulation of spinodal decomposition took place in the canonical 
ensemble. Again, a homogeneous starting configuration was equilibrated
at a higher temperature and quenched to a state below a
critical point of the mixture system. We only allowed for local 
Monte Carlo displacements which should yield kinetics comparable 
to those obtained in Molecular Dynamics.

\section{Static equilibrium properties}

\begin{figure}
\begin{center}
\epsfig{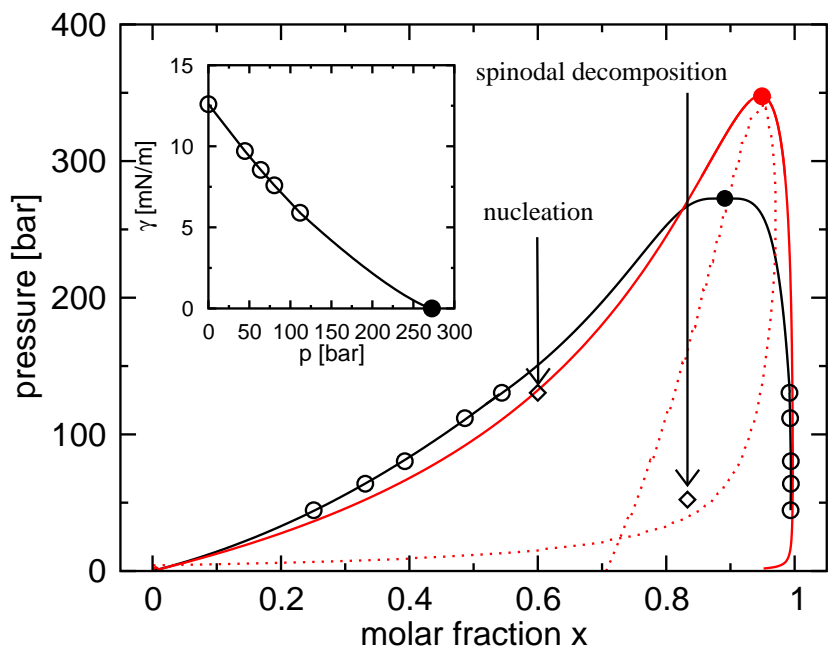}
\caption{\label{fig4} Isothermal slice through the
phase diagram in Fig.~\ref{fig3} at temperature T=486 K. Black curve with circles
shows the simulated coexistence curve. Red 
curves are corresponding predictions from TPT1 theory \cite{23}
for coexistence (full) and the spinodal curve (dotted), respectively.
The position of the critical points are denoted by $\bullet$.
$\diamond$ indicate phase diagram positions of nucleation and spinodal
decomposition movies in Fig.~\ref{fig6} and Fig.~\ref{fig7}. 
(Inset): surface tension as a function of pressure.
}
\end{center}
\end{figure}

First we return to the phase diagrams in
Figs.~\ref{fig3},~\ref{fig4} and discuss their properties more
closely. Note that at the temperature chosen in Fig.~\ref{fig4}
($T=486$~K) we are very far below the critical point of pure
hexadecane (which is $T_c=723$ K). Therefore the hexadecane melt
(with some dissolved carbon dioxide) coexists with almost pure
CO$_2$ the composition of which gradually changes from gas- to liquid-like with
increasing pressure. In both cases there is almost no hexadecane present.
This is reflected in the coexistence curve which rapidly
approaches the CO$_2$ molar fraction $x=1$.

We have also included the results of an
analytical approximation for the phase diagram of our model,
obtained from a thermodynamic perturbation theory \cite{23} called
TPT1, developed along the lines of Wertheim et al.~\cite{27}. It
is seen that this theory describes the coexistence curve in Fig.\ref{fig4} at
pressures that are much smaller than the critical pressure
$p_{\textrm{crit}} (T)$ almost quantitatively, while the critical
pressure itself is clearly overestimated. Of course, such an
overestimation of the critical parameters $p_{\textrm{crit}} (T)$,
$x_{\textrm{crit}} (T)$ is quite typical for all theories that
invoke a mean field description of the critical behavior, as TPT1
does. A similar analytical study for more alcane-CO$_{2}$ systems has been
performed recently \cite{blas}.
Note also that the TPT1 theory readily yields a spinodal
curve, which has the standard
meaning of separating ``metastable states'' from ``unstable
states'' in the phase diagram \cite{28,29,30}. The naive (mean-field) description
implies that in the metastable region of the phase diagram the
initial stages of phase separation kinetics are described by
nucleation and growth \cite{28,29}, while in the unstable region
the decay mechanism is spinodal decomposition \cite{29,30}. In 
systems with short range forces
there is no well-defined sharp spinodal line, the actual
transition between both decay mechanisms occurs rather gradually
in a broad transition region, and this region is not centered
around the mean field spindodal but occurs close to the
coexistence curve \cite{30}. However, for the quenching experiment
performed at $T=486$~K and $x=0.60$, where the final state
corresponds to a pressure of about $p\approx$130~bar, 
we are so close to the coexistence curve and so far
from the mean field spinodal, that classical nucleation-type
behavior \cite{28,29} should be observable. In order
to demonstrate the mechanism of spinodal decomposition, we quenched
into the unstable region as defined by the mean-field spinodal
(compare with Fig.~\ref{fig4}).

\begin{figure}
\epsfig{file=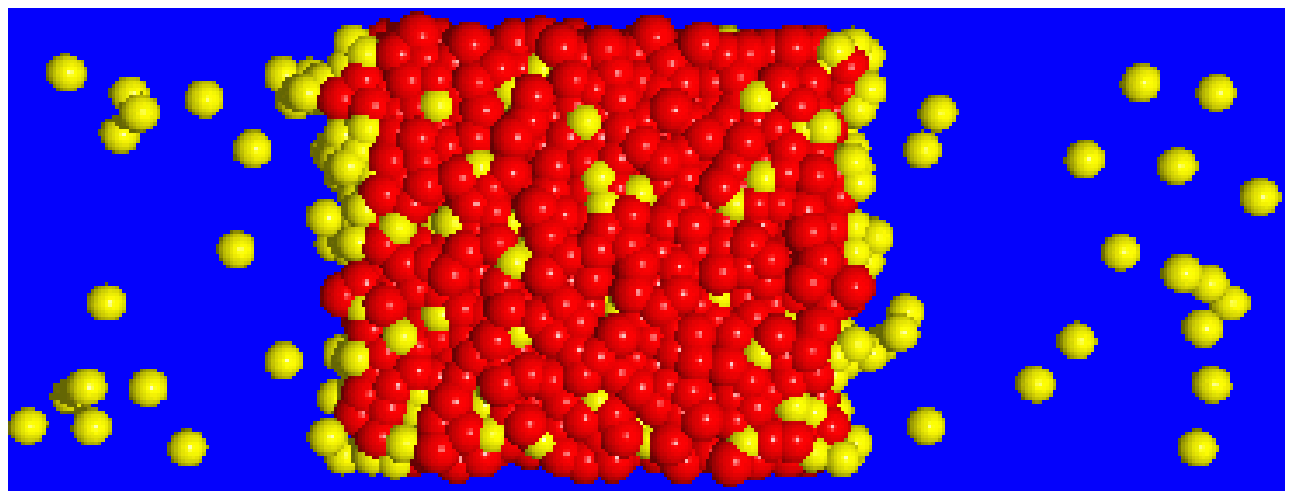,width=1.0\linewidth,clip=}
\epsfig{file=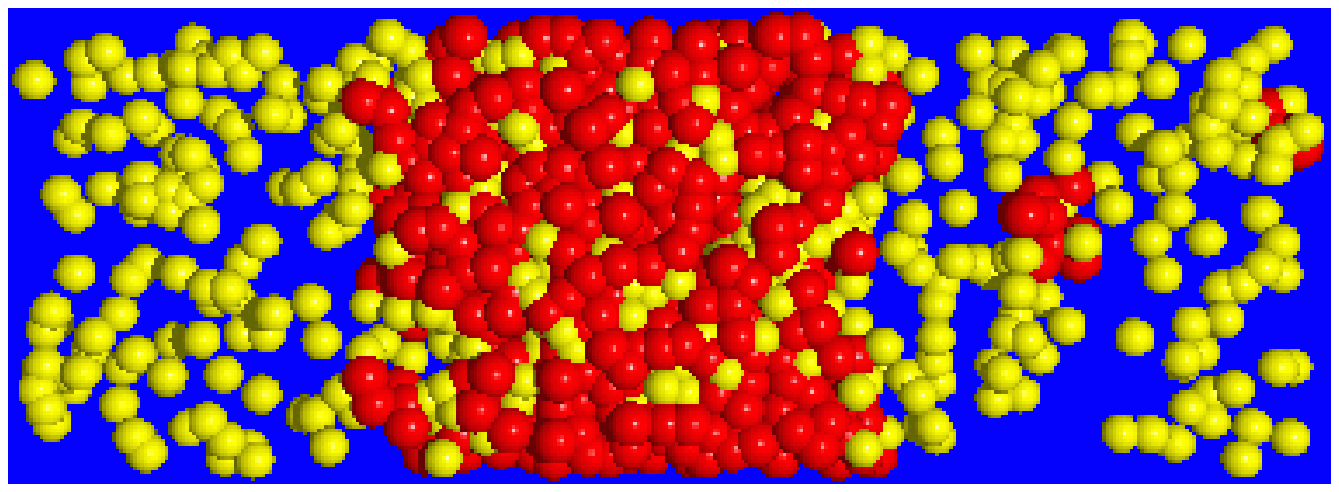,width=1.0\linewidth,clip=}
\caption{\label{fig5} Snapshot pictures of systems
in the center of the two-phase coexistence region for T=243~K and T=486~K. 
Snapshots show a 2$\sigma_{hh}$ slice through a box with
dimensions $L_x \times L_y \times L_z = $18 $\sigma_{hh}$ $\times$ 18 $\sigma_{hh}$ $\times$ 54 $\sigma_{hh}$. 
Positions of particles are projected into the $xz$ plane. CO$_2$ particles are
shown as yellow circles of radius $\sigma_{cc}$, while effective monomers of
hexadecane are shown as red circles of radius $\sigma
_\textrm{hh}$ (we choose $\sigma _{\textrm{hh}} =1$ as our unit of
length, which then implies $\sigma_{\textrm{cc}}=0.816$),
the background is blue. For T=243~K interfacial wetting can be observed.}
\end{figure}

Classical nucleation theory assumes (in our case) the formation of
spherical gas bubbles of essentially pure CO$_2$ within a hexadecane 
matrix . For a quantitative
understanding of the free energy barrier against homogeneous
nucleation, clearly the understanding of the
interface between macroscopic coexisting
phases is a prerequisite. As discussed elsewhere \cite{17,18},
such interfaces are most conveniently studied in the
grandcanonical ensemble in conjunction with a multicanonical
preweighting scheme that generates mixed-phase configurations.
Choosing a simulation box of rectangular shape $L \times L \times
cL$ with a constant $c=2$ or larger and periodic
boundary conditions, one generates system configurations where the
two coexisting phases are separated by two parallel $L \times L$
interfaces (compare with Fig.~\ref{fig5}a). ``Measuring'' the canonical probability of these
two-phase configurations relative to the probability of the pure
phases is a standard method for the estimation of interfacial free
energies \cite{16,17,18,31,32,33,34,35,36}. The results for T=486 K are
shown in the inlet of Fig.~\ref{fig4}.
In addition, this method can be used to generate well-equilibrated configurations of
interfaces allowing a study of the interfacial structure.
Fig.~\ref{fig5} shows, as an example, two snapshot pictures of
such states containing interfaces, one at $T$=486~K and the other at $T$=243~K. 
The size of  the particles is
enlarged (radii shown are $\sigma_{\textrm{cc}}$,
$\sigma_{\textrm{hh}}$ rather than $\sigma_{\textrm{cc}}/2$,
$\sigma_{\textrm{hh}}/2$), for the sake of a clearer view. The
temperature $T$=243~K is
slightly lower than the temperature of the critical point of pure
CO$_2$, and thus the system
possesses a triple point where a polymer-rich phase coexists with
two CO$_2$-rich phases, one being liquid CO$_2$ and the other
the gas. Consequently, when we study the interface between the
polymer-rich phase and gaseous CO$_2$, a layer of the third phase
(liquid CO$_2$) intrudes at this interface. The thickness of this
layer is expected to diverge when the pressure approaches the
value of the triple point pressure. For higher pressures, we then
have a coexistence between the polymer rich-phase and the
liquid-like dense CO$_2$, similar to the situation that occurs at
the higher temperature (Fig.~\ref{fig5}a). 
At low enough temperatures, where a triple point
occurs, one expects a strong decrease of the interfacial tension
already when the triple point pressure is approached \cite{11}:
the interfacial tension between the polymer-rich phase and
liquid-like CO$_2$ is much smaller than the interfacial tension
between the polymer-rich phase and the gas.
The barrier against nucleation also decreases
when one approaches the spinodal from the metastable region.
This fact facilitates nucleation of relatively large droplets observable on the
time-scales of our Monte Carlo simulation.

\section{Monte Carlo simulation of bubble nucleation}

\begin{figure}
\begin{center}
\epsfig{file=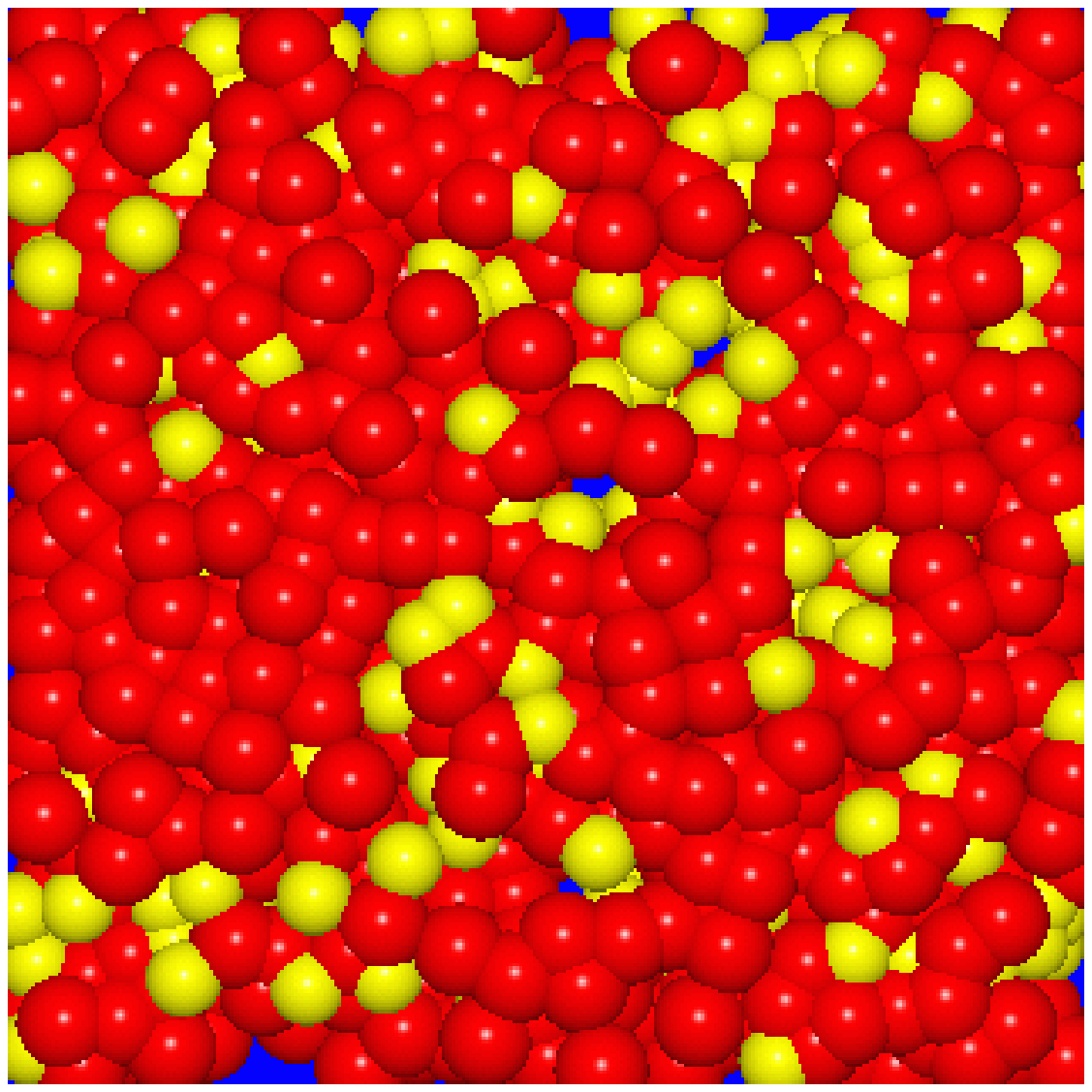,width=0.49\linewidth,clip=}
\epsfig{file=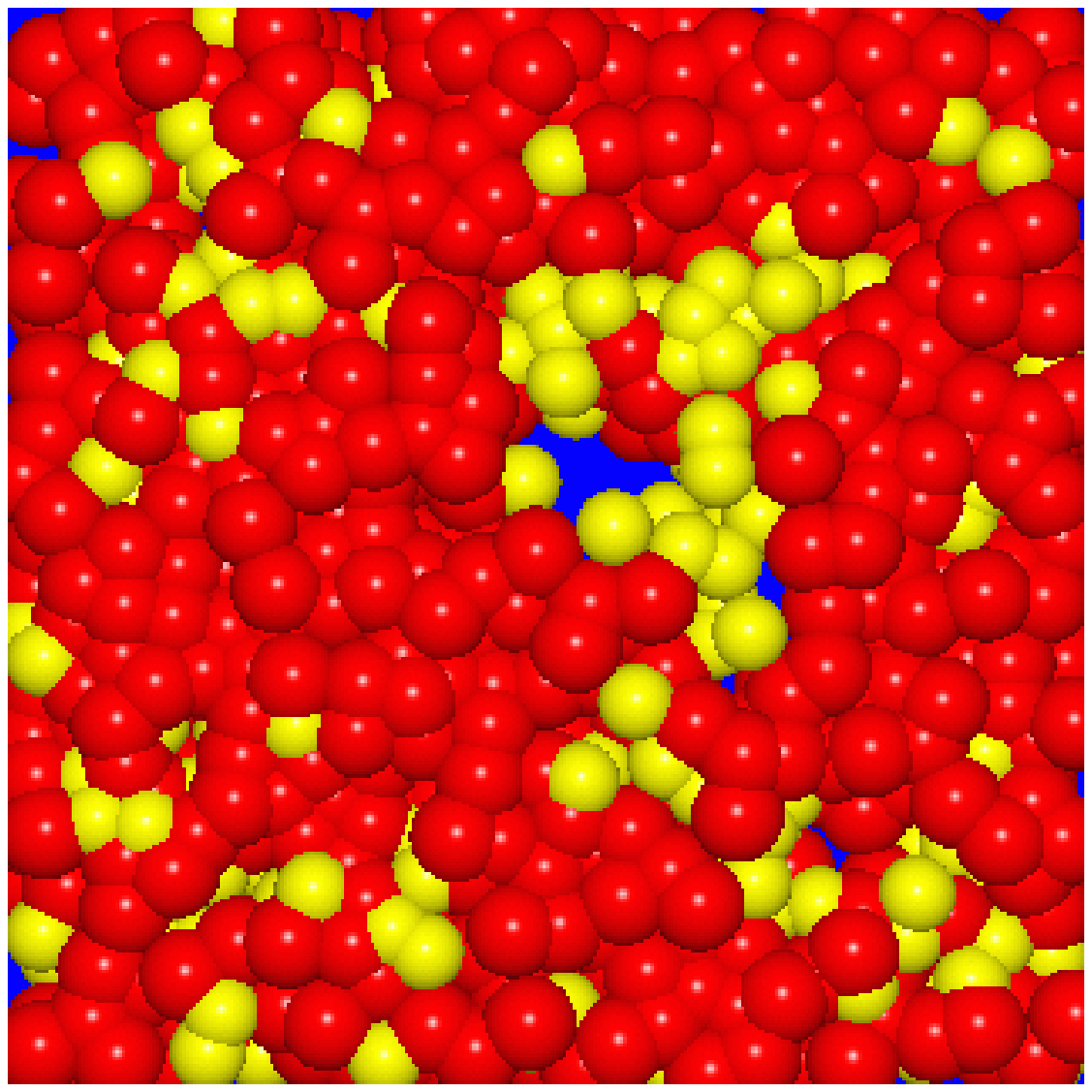,width=0.49\linewidth,clip=}
\epsfig{file=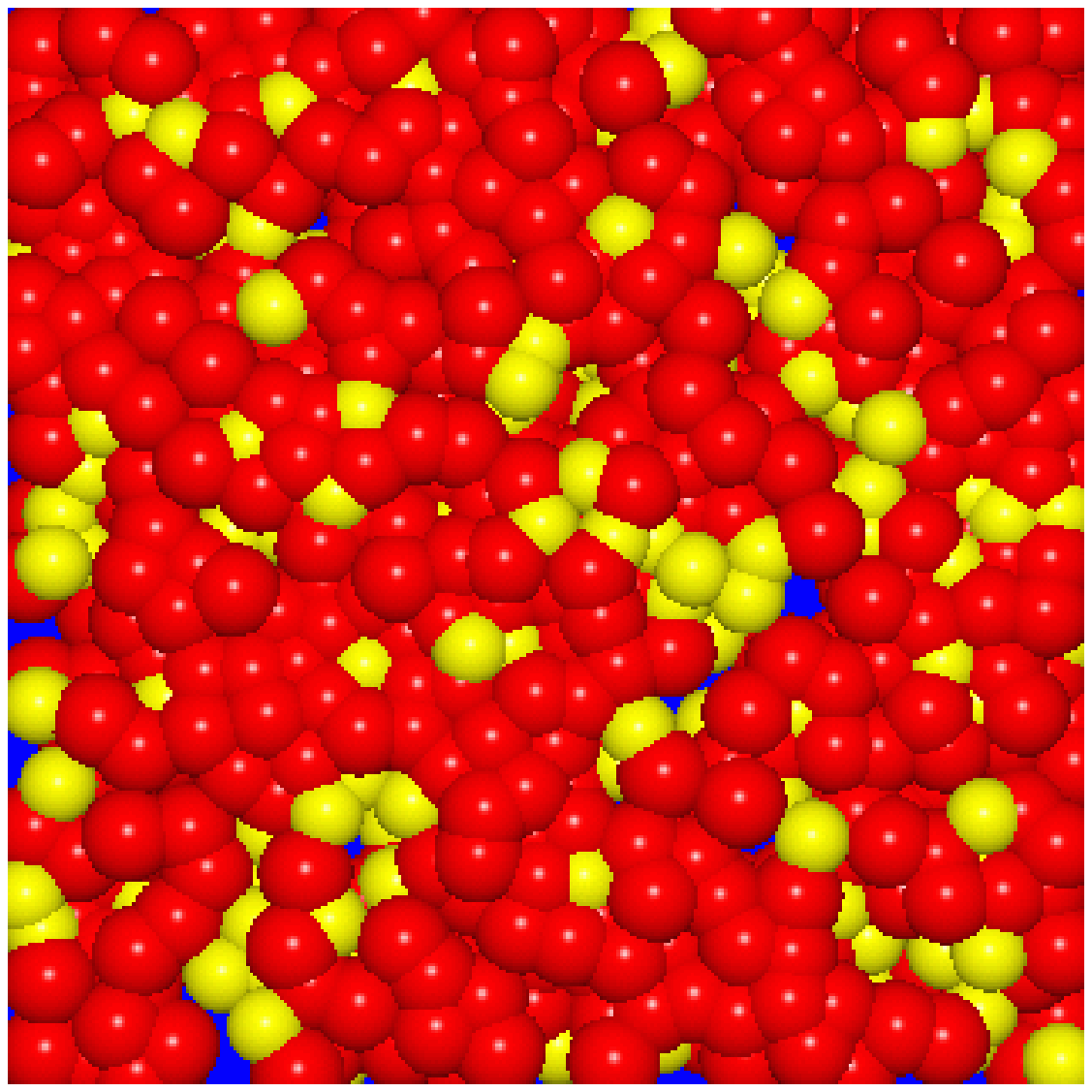,width=0.49\linewidth,clip=}
\epsfig{file=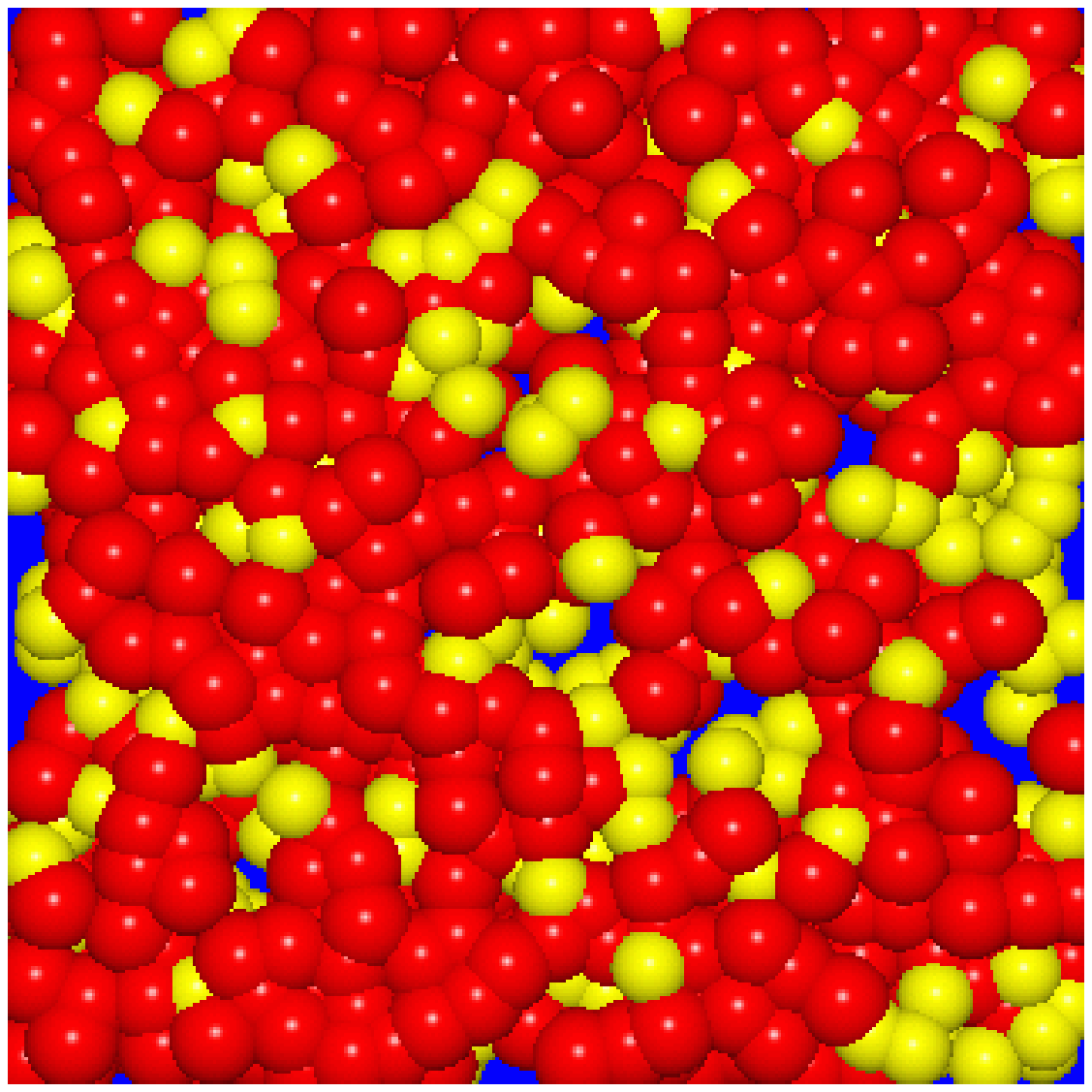,width=0.49\linewidth,clip=}
\epsfig{file=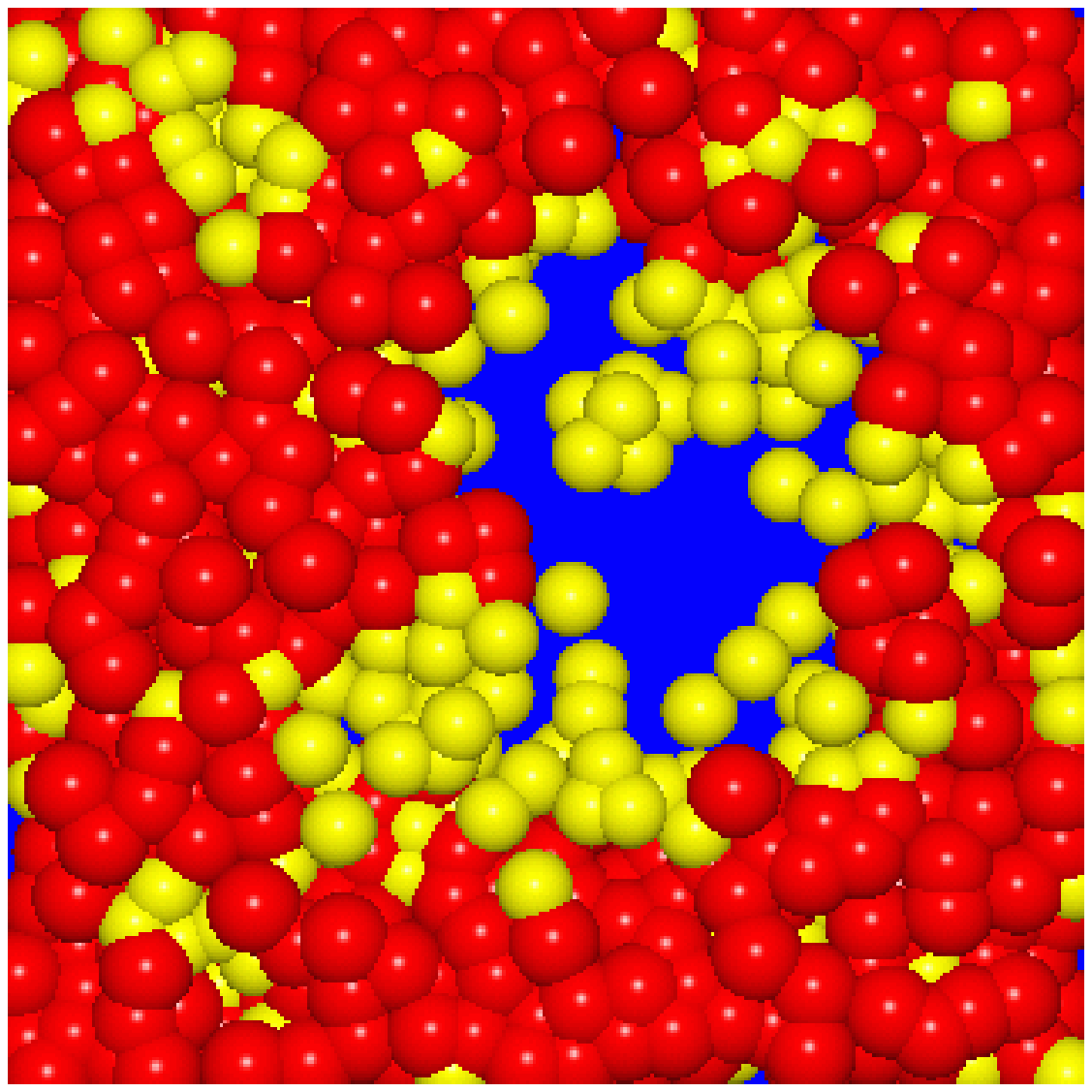,width=0.49\linewidth,clip=}
\epsfig{file=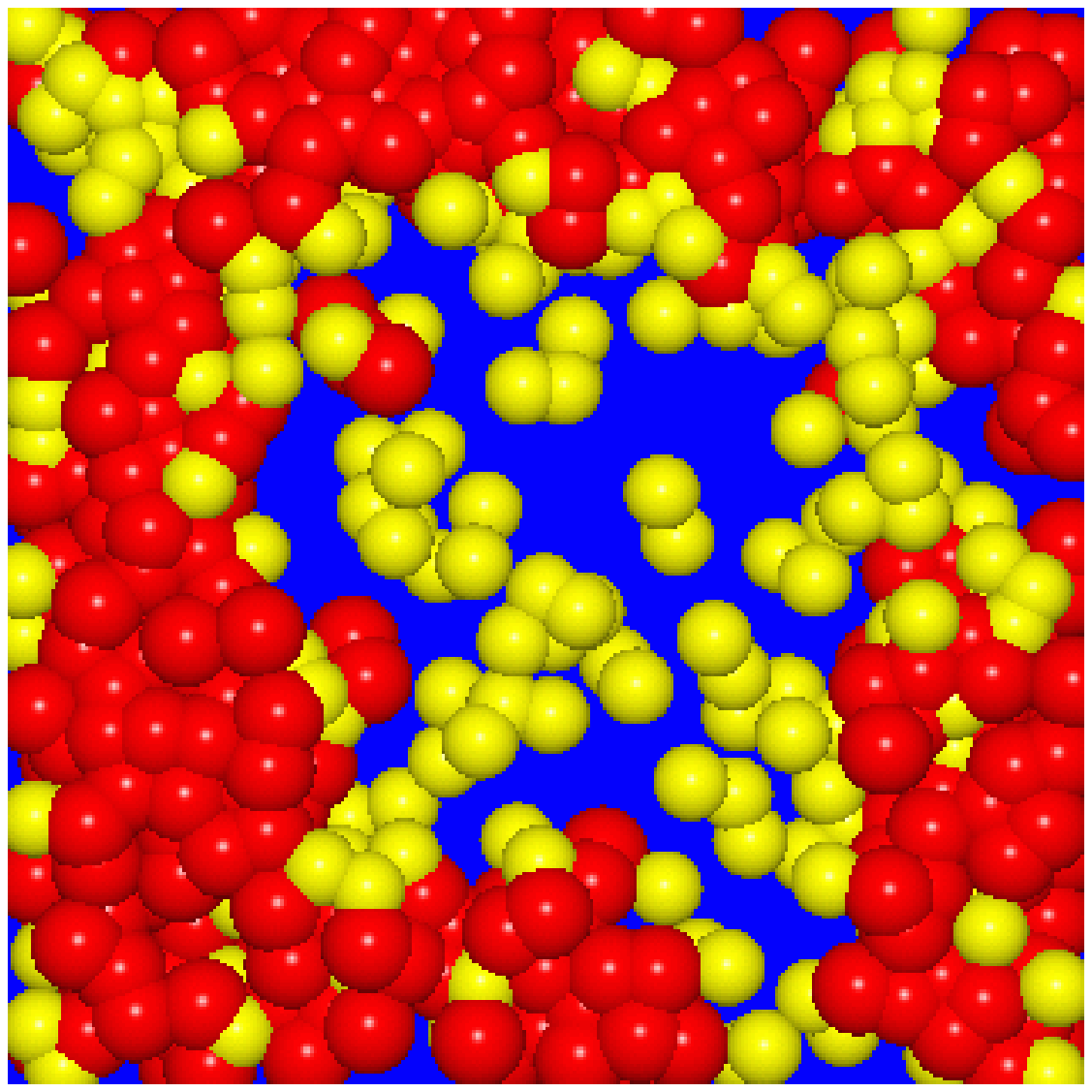,width=0.49\linewidth,clip=}
\caption{\label{fig6} (Remark: The final version (submitted to New Journal of Physics) will contain a short
animation. Here, only snap-shot pictures are shown.)
A short animation showing the time evolution of a quenching experiment
into the metastable region ($T$=486~K, $x$=0.60 and (final) pressure $p \approx$ 130 bar 
- compare with Fig.\ref{fig4}). Phase separation occurs via nucleation.
The linear dimension of the simulation box is 22.5~$\sigma_{hh}$.
The movie is generated from subsequent snapshot pictures of
a grandcanonical Monte Carlo simulation. For clarity, not the whole simulation but
only a slice of thickness 2~$\sigma_{hh}$ is shown.
The color coding is the same as in Fig.~\ref{fig5}.}
\end{center}
\end{figure}

The animation in Fig.~\ref{fig6} visualizes how phase separation
in polymer solutions proceeds via nucleation. In the beginning
the system fluctuates around a meta\-stable free energy minimum.
One observes the ``birth and death'' of small density fluctuations
in the hexadecane matrix (displayed as red spheres)
which become visible whenever the blue background shines through
the slice. These irregular voids are too small, however, to lead
to an immediate decay of the initial metastable state. Usually
the voids also contain a few CO$_2$ molecules (displayed as small yellow spheres). 
Only after some time lag a void manages to grow to critical size
and overcome the free energy barrier which seperates the metastable
from the homogeneous equilibrium state. From now on the bubble 
will grow until it fills the whole simulation box.
As expected, the critical void gets also filled with CO$_2$ molecules,
thus decreasing the supersaturation of the remaining polymer-rich phase.
Remarkably, this filling of the bubble does not occur
homogeneously: at the interface between the gas and the polymer,
i.e., at the surface of the bubble, the CO$_2$ density is clearly
higher than it is in the center of the bubble. This surface
enrichment effect may be interpreted as a precursor to interfacial wetting 
of the fluid CO$_2$ predicted when one approaches the triple point
\cite{11}. A quantitative analysis of this phenomena is in preparation \cite{yucheng}. 

Of course, the observation of a nucleation event displayed in
Fig.~\ref{fig6} is by no means the first
observation of nucleation phenomena by computer simulations: e.g.,
already Ref. \cite{28} contains a series of snapshot pictures
illustrating nucleation in the two-dimensional lattice gas model.
Since then many more elaborate studies have appeared. However,
the present work deals with nucleation in a system of industrial
relevance, namely bubble nucleation in metastable polymer melts
supersaturated with supercritical CO$_2$. In addition, the
nucleation mechanism here is quite nontrivial, since the
``critical bubble'' (which just has the size to be at the
nucleation barrier, where there is a 50\% chance of growth or decay) is
characterized by two variables, its size and the number of CO$_2$
molecules it contains. Similarly the interface is described by
nontrivial profiles of two variables, the total density and the
relative concentration of CO$_2$. Such a nontrivial coupling
between two order parameters has also been predicted by
recent self-consistent field calculations based on a qualitatively
similar model \cite{11}. Since
these mean field descriptions do not take into account the
effects of fluctuations, a check by computer simulation methods is clearly warranted.

\begin{figure}
\begin{center}
\epsfig{file=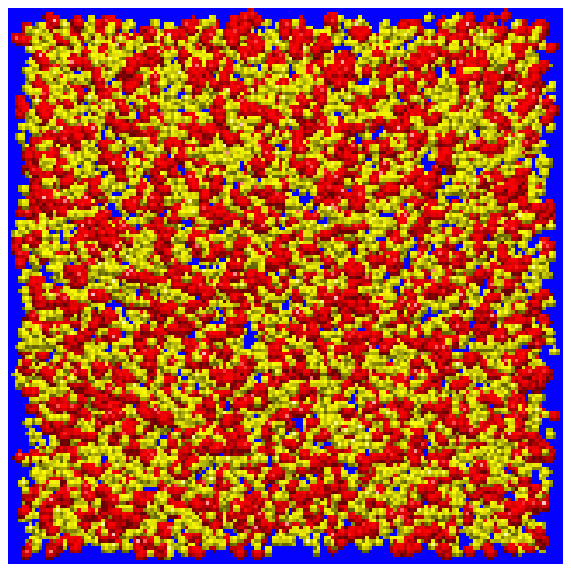,width=0.32\linewidth,clip=}
\epsfig{file=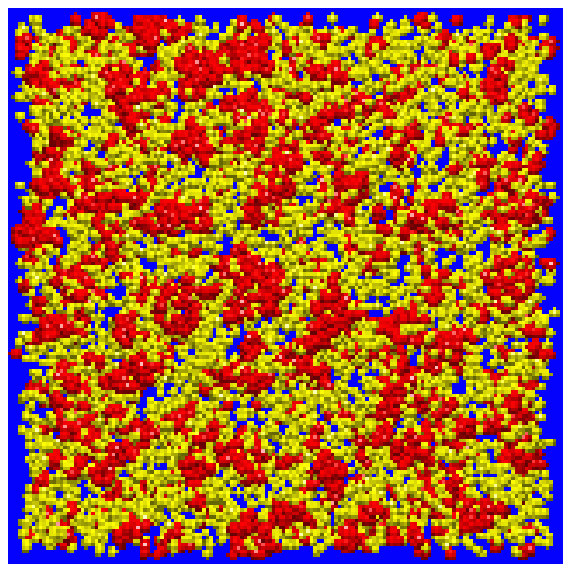,width=0.32\linewidth,clip=}
\epsfig{file=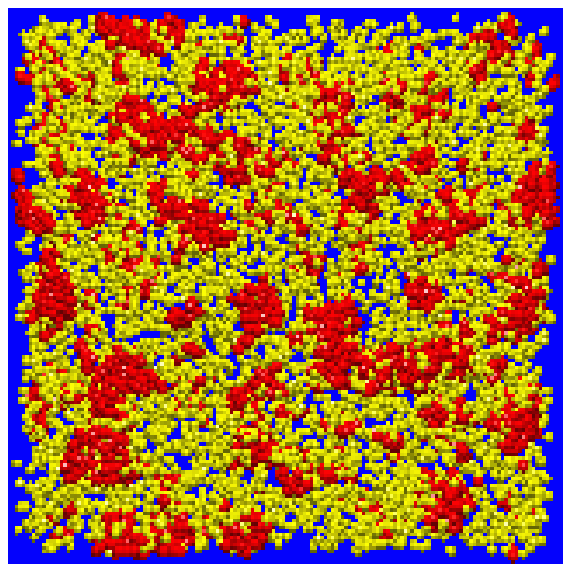,width=0.32\linewidth,clip=}
\caption{\label{fig7} A series of snapshots visualizing the early stages of spinodal
decomposition after a quenching experiment into the unstable region ($T$=486 K, $x$=0.833 
and (final) pressure $p \approx$ 52 bar 
- compare with Fig.~\ref{fig4}). Phase separation occurs via spinodal decomposition.
The size of the box is 91.21 $\sigma_{hh}$ $\times$ 91.21 $\sigma_{hh}$ $\times$ 17.53 $\sigma_{hh}$,
but only a slice of 10 $\sigma_{hh}$ is shown. The system is actually more dilute than the one shown in
Fig.~\ref{fig6}. 
(a) Starting configuration, (b) after 125000 and (c) after 500000 local displacements (0-0.3 $\sigma_{5}$)
per particle. The color coding is the same as in Fig.~\ref{fig5}.}
\end{center}
\end{figure}

When we quench to a position below the critical point (Fig.~\ref{fig7}), the
behavior is rather different: fluctuations in
the relative concentration of CO$_2$ get gradually more pronounced
everywhere. These fluctuations are not localized, however, in the form
of identifiable bubbles, but rather form an irregular percolating
network. This is the hallmark of spinodal decomposition
\cite{29,30}. Since the linear theory of spinodal
decomposition only holds for systems
with long range forces in the very early stages \cite{30}, we do
not attempt a more quantitative analysis of this data. Note also
that the build-up of these concentration fluctuations is
relatively faster than the nucleation of bubbles. Hence the
separation of time scales between the structural relaxation and
the rate constant of spinodal decomposition is less well
established. In the intermediate stages of spinodal decomposition,
we also expect hydrodynamic mechanisms to prevail \cite{29,30},
which are not captured by Monte Carlo moves.

\begin{figure}
\begin{center}
\epsfig{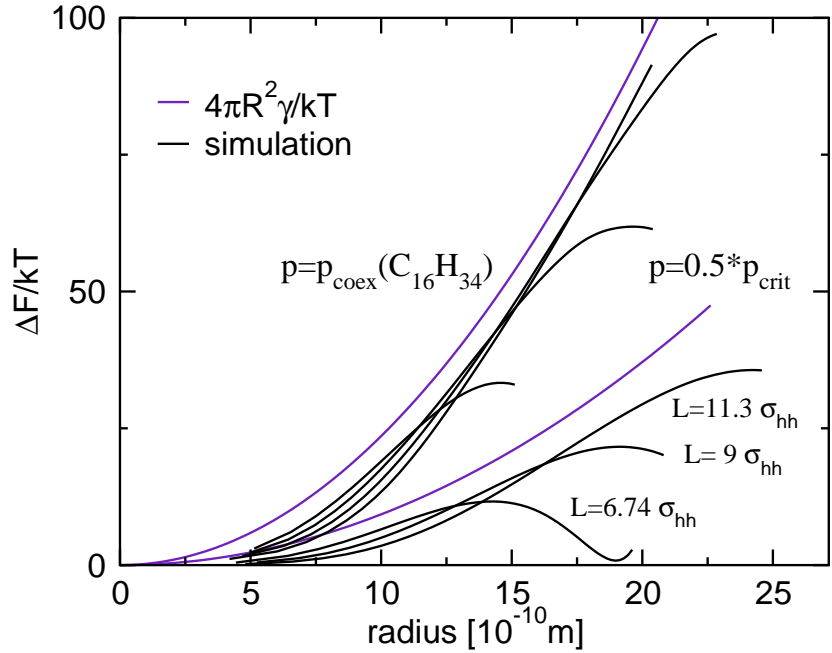}
\caption{\label{fig8}Free energy as a function of droplet size for T=486 K and $p=0.5~p_{crit}$ and
p$\approx$0 (coexistence pressure of pure hexadecane - compare with Fig.~\ref{fig3} and Fig.~\ref{fig4}). 
Indigo line: A simple estimate for the free energy: $\Delta F=4\pi\gamma R^{2}$,
surface tension $\gamma$ is taken from Fig.~\ref{fig4} (inlet) (flat surface). Black: results from simulations of
different system sizes. Only the envelope of the curves is relevant. Other parts of the
curves belong to regions of the distribution where no nucleation is expected.
For small droplets the free energy
is smaller than expected. Differences decrease with increasing radius and decreasing distance from
the critical point of the isotherm.
}
\end{center}
\end{figure}

Returning now to the simulation of nucleation phenomena, we ask
how one can quantify these observations. A straightforward 
but tedious method would
be to repeat dynamic simulations as shown in Fig.~\ref{fig6} many
times in order to extract an estimate of the nucleation rate from
the average time it takes to form a bubble or a droplet. Of course, since the
nucleation rate is expected to decrease dramatically if the quench
depth is decreased slightly, such a direct method works hardly in
practice. An indirect method to obtain information on the surface
free energy of clusters as a function of cluster size simply uses
the final equilibrium states of simulations. An estimate
for the free energy as a function of density can be obtained directly
from the probability distribution. In this context special care has to be taken
of finite-size effects to identify regions of densities which correspond to a
single cluster \cite{nucl1,nucl2}.
If we assume that bubbles (or droplets) are spherical,
filled with gas and surrounded by a homogeneous liquid of coexistence
density (or vice versa), we can assign a radius to each density and hence a
free energy estimate to each radius. 
Fig.~\ref{fig8} indicates that the
surface free energy obtained this way
is somewhat smaller than expected from the so-called 
``capillarity approximation''. A simple 
estimate of the free energy is $F_s=4\pi R^2\gamma_{\rm int}$,
where $R$ is the radius of the cluster and $\gamma_{\rm int}$ the
interfacial tension of flat interfaces as shown in
Fig.~\ref{fig5} and derived in Fig.~\ref{fig4}. As a
consequence, the nucleation barrier is lower than the
classical theory of homogeneous nucleation \cite{28,29} would
predict. Differences in $\gamma_{\rm int}$ also decrease with increasing
cluster size and distance to the critical point of the mixture. At this point, however, the findings are still somewhat
preliminary, and more work will be necessary to obtain
sufficiently accurate results \cite{39}.

\section{Concluding Remarks}

In summary, we have presented a simple model and computational tools which allow us to examine equilibrium
properties and kinetics of phase separation in systems of industrial relevance.
We were able to produce a two-component mixture phase diagram with a complete critical line which
is in qualitative agreement with experiments. For a system at T=486 K we determined
the isotherm and interfacial tension of a flat surface as a function of pressure.
In a small animations and a series of snapshots we visualized bubble nucleation and
spinodal decomposition for this specific system. For the first case we found a precursor of
interfacial wetting which was also observed on a flat surface in the vicinity 
of the three-phase coexistence line. Finally, we found evidence that for small bubbles the interfacial
tension is smaller than expected by classical theory of homogeneous nucleation. 

\section{Acknowledgements}

Financial support from the BASF AG (P.V.) and from the Deutsche Forschungsgemeinschaft (DFG) via a
Heisenberg fellowship (M.M.) and by grant No BT 314/17-3 (L.G.M) is gratefully acknowledged.
LGM would like to thank Ministerio de
Ciencia y Tecnologia (MCYT) and Universidad Complutense 
(UCM) for the award of a Ramon y Cajal fellowship and for
financial support under contract BFM-2001-1420-C02-01.
We are grateful to E. H\"adicke, B. Rathke, R. Strey, and D.W. Heermann for stimulating discussions, and to 
NIC J\"ulich, HLR Stuttgart, and ZDV Mainz for generous grants of computer time. 

\newpage
\pagestyle{empty}

\end{document}